\documentclass[preprint,12pt]{elsarticle}

\usepackage[right]{lineno}
\usepackage{amssymb}
\usepackage{amsthm}
\usepackage[version=4]{mhchem}
\usepackage{lscape}
\usepackage{hyperref}
\usepackage{color}
\hypersetup{colorlinks=true}

\journal{Journal of Volcanology and Geothermal Research}

\begin{document}

\begin{frontmatter}



\title{Surface tension estimation of bubble nuclei in magma using spinodal pressure and nonclassical nucleation theory}

\author[inst1]{Mizuki Nishiwaki\corref{cor}}
\ead{nishiwaki.m(at)mat.usp.ac.jp}

\affiliation[inst1]{organization={Center for Glass Science \& Technology, School of Engineering, The University of Shiga Prefecture},
            addressline={2500, Hassaka-cho}, 
            city={Hikone, Shiga},
            postcode={522-8533}, 
            country={Japan}}

\begin{abstract} 
Efforts to estimate the magma decompression rate from the vesicular texture of volcanic products have progressed through the development of theoretical models and laboratory experiments.
The theoretical model is based on nucleation theory, with the surface tension between the melt and bubble nucleus being the parameter that most strongly governs nucleation.
Since direct measurement of surface tension is difficult, it has been calculated by inverting the bubble number density from experimental samples using classical or nonclassical nucleation theory formulas.
However, in the nonclassical case, which accounts for the supersaturation dependence of surface tension, the pressure at the spinodal limit (where surface tension becomes zero) was previously unknown, necessitating complex mathematical operations.
In this study, the spinodal pressure determined from the Gibbs energy curve was substituted into the nonclassical formula by approximating the water-saturated silicate melt as a two-component symmetric regular solution composed of silicate and water.
This approach allowed for a more straightforward estimation of surface tension using data from past decompression experiments.
Nevertheless, the resulting surface tension values were more scattered than those obtained using the classical formula, suggesting that applying the nonclassical formula to magma vesiculation is not valid at present. 
Resolving this issue will likely require an integrated understanding of the dependence of surface tension on both supersaturation and bubble radius.
Such understanding would enable more accurate estimation of surface tension and contribute to reconciling the discrepancy between theoretical and experimental bubble number density values.
\end{abstract}

\begin{highlights} 
\item Supersaturation dependence of the bubble nucleus' surface tension was revisited. 
\item Surface tension estimation by nonclassical nucleation theory remains challenging. 
\item Surface tension dependence on supersaturation and bubble size needs integration. 
\end{highlights}


\end{frontmatter}

\begin{table}[h]
\caption{Notation list.}
\begin{flushleft}
\begin{tabular}{lll} \hline
Symbol & Unit & Definition \\ \hline
$a_0$ & m & Average distance between water molecules in the melt \\ 
$D_\mathrm{\ce{H2O}}$ & m$^2$ s$^{-1}$ & Diffusivity of total water in the melt \\ 
$f$ & Pa & Fugacity \\
$J$ & No m$^{-3}$ s$^{-1}$ & Nucleation rate \\ 
$k_\mathrm{B}$ & J K$^{-1}$ & Boltzmann's constant \\ 
$n_0$ & No m$^{-3}$ & Number of water molecules per unit melt volume \\ 
$P$ & Pa & Pressure \\
$P_\mathrm{bi}$ & Pa & Pressure on the binodal curve (= binodal pressure) \\
$P_\mathrm{spi}$ & Pa & Pressure on the spinodal curve (= spinodal pressure) \\
$P_\mathrm{M}$ & Pa & Melt pressure \\
$P_\mathrm{SAT}$ & Pa & Water saturation pressure \\
$P^*_\mathrm{B}$ & Pa & Internal pressure of the critical bubble nucleus \\ 
$R$ & m & Radius of a bubble nucleus \\
$R_\mathrm{c}$ & m & Critical bubble radius \\
$T$ & K & Temperature \\
$\overline{V}_\mathrm{\ce{H2O}}$ & m$^3$ mol$^{-1}$ & Partial molar volume of water in the melt \\ 
$x$ & no unit & Mole fraction of one of the two components \\
$\delta_\mathrm{T}$ & m & Tolman length \\
$\xi$ & No & Index of supersaturation (overpressure) defined by Eq. (\ref{xi}) \\
$\sigma$ & N m$^{-1}$ & 
\begin{tabular}{l} 
``Microscopic'' surface tension between the melt and \\ homogeneous spherical bubble nucleus with a large curvature
\end{tabular} \\
$\sigma_\mathrm{GG}$ & N m$^{-1}$ & 
\begin{tabular}{l} 
``Microscopic'' surface tension evaluated in Gonnermann \\ and Gardner (2013) by their method
\end{tabular} \\
$\sigma_\mathrm{Shea}$ & N m$^{-1}$ & 
\begin{tabular}{l} 
``Microscopic'' surface tension calculated by inverting the \\ bubble number density (BND) of decompression experiments \\ using the classical nucleation theory (CNT) formula in Shea (2017) 
\end{tabular} \\
$\sigma_\infty$ & N m$^{-1}$ & 
\begin{tabular}{l} 
``Macroscopic'' surface tension at the flat interface between \\ the melt and vapor 
\end{tabular} \\ \hline
\end{tabular}
\end{flushleft}
\label{Table1}
\end{table}

\section{Introduction}
\label{Introduction}

The vesicular texture observed in volcanic products has long been considered to record the dynamics within the conduit from the magma chamber to the ground surface. For example, elementary processes such as bubble coalescence, elongation, and rupture have been investigated using various approaches, including the analysis of natural samples, theoretical modeling, and laboratory experiments (e.g., Ohashi et al., 2021; Jones et al., 2022; Maruishi and Toramaru, 2024).
In particular, the bubble number density (BND)---the number of bubbles per unit volume---has been shown to strongly correlate with the decompression rate of magma, as demonstrated by both theoretical models and laboratory experiments. Based on classical nucleation theory (CNT), Toramaru (2006) developed a practical tool called the BND decompression rate meter, which allows the estimation of decompression rates using BND along with specific values for the following parameters: temperature $T$, initial water saturation pressure $P_\mathrm{SAT}$, water diffusivity in the melt $D_\mathrm{\ce{H2O}}$, and the “microscopic” surface tension $\sigma$ between the melt and a homogeneous spherical bubble nucleus.

The parameters $T$ and $P_\mathrm{SAT}$ can be controlled in experimental systems, and also in natural systems, they can be estimated based on previous experimental petrology data (e.g., water solubility or geothermobarometry using mineral assemblages). $D_\mathrm{\ce{H2O}}$ is also well studied experimentally for its temperature, pressure, and water content dependence, making it relatively easy to compute (e.g., Zhang and Ni, 2010). These parameters, even with some uncertainty in their estimates, have only a minor impact on the resulting nucleation and decompression rates (Shea, 2017).
In contrast, the estimated value of $\sigma$ significantly affects the calculated nucleation and decompression rates (Shea, 2017), highlighting that $\sigma$ is the most influential parameter in governing nucleation. Constraining $\sigma$ is key for achieving consistency between the results of magma decompression experiments and CNT, and for further improving the ability of BND decompression rate meter to accurately estimate the decompression rate of natural magmas.
However, due to the difficulty in directly measuring $\sigma$, Toramaru (2006) adopted the capillary approximation, assuming that $\sigma$ is equal to the “macroscopic” surface tension $\sigma_\infty$ of a flat melt interface. The functional form of $\sigma_\infty$ as a function of $T$ and $P_\mathrm{SAT}$ was taken from the measurements by Bagdassarov et al. (2000).

An alternative method for estimating $\sigma$ is by fitting the integrated CNT-based nucleation rate $J$ over decompression time to the BND obtained from decompression experiments (i.e., inverting BND using the CNT formula) (e.g., Mourtada-Bonnefoi and Laporte, 2004; Cluzel et al., 2008; Hamada et al., 2010; Shea, 2017). See \ref{Appendix} for the detailed expression of $J$ in CNT (e.g., Hirth et al., 1970).
Contrary to the capillary approximation, the inversion method yielded a lower $\sigma$ than $\sigma_\infty$. 
Gonnermann and Gardner (2013) attributed this discrepancy to applying the capillary approximation, which is strictly valid only near equilibrium, to a non-equilibrium melt--bubble nucleus system with large supersaturation. According to the recent nonclassical nucleation theory (non-CNT), the interface between the original and new phases loses sharpness and diffuses under non-equilibrium conditions (e.g., Chapter 4 in Kelton and Greer, 2010). In other words, such interfacial diffusion should also occur during the vesiculation of magma with supersaturated water, which is no longer soluble due to decompression.
Based on the non-CNT, they considered a situation where $\sigma$ depends on the degree of supersaturation: the greater the degree of supersaturation, the smaller $\sigma$ becomes, and this situation was expressed mathematically, as described below.
However, the pressure at the “spinodal limit,” where the system is far from equilibrium and $\sigma = 0$, referred to as the spinodal pressure $P_\mathrm{spi}$, was an unknown parameter at that time.

Recently, Nishiwaki (2025) investigated the possibility of spinodal decomposition occurring during decompression-induced magma vesiculation, as Allabar and Nowak (2018) proposed, from a thermodynamic perspective. He treated hydrous magma as a symmetric regular solution consisting of two components: an anhydrous silicate melt and water vapor. Excess and mixing Gibbs energies were determined, and based on the shape of these energies, simple calculations were performed to estimate the binodal pressure $P_\mathrm{bi}$ ($= P_\mathrm{SAT}$) and spinodal pressure $P_\mathrm{spi}$ at a certain mole fraction of water.
By substituting the value of $P_\mathrm{spi}$ into the non-CNT model by Gonnermann and Gardner (2013), it is anticipated that the value of $\sigma$ can be determined more straightforwardly. This study aims to explore this method and assess the validity of their model’s application to magma vesiculation. Based on these results, I present my personal perspective on the future direction of microscopic surface tension research.

\section{Supersaturation dependence of the surface tension: reinterpretation of Gonnermann and Gardner's (2013) idea}
\label{GG13}

\begin{figure}
    \centering
    \includegraphics[width=1.1\linewidth]{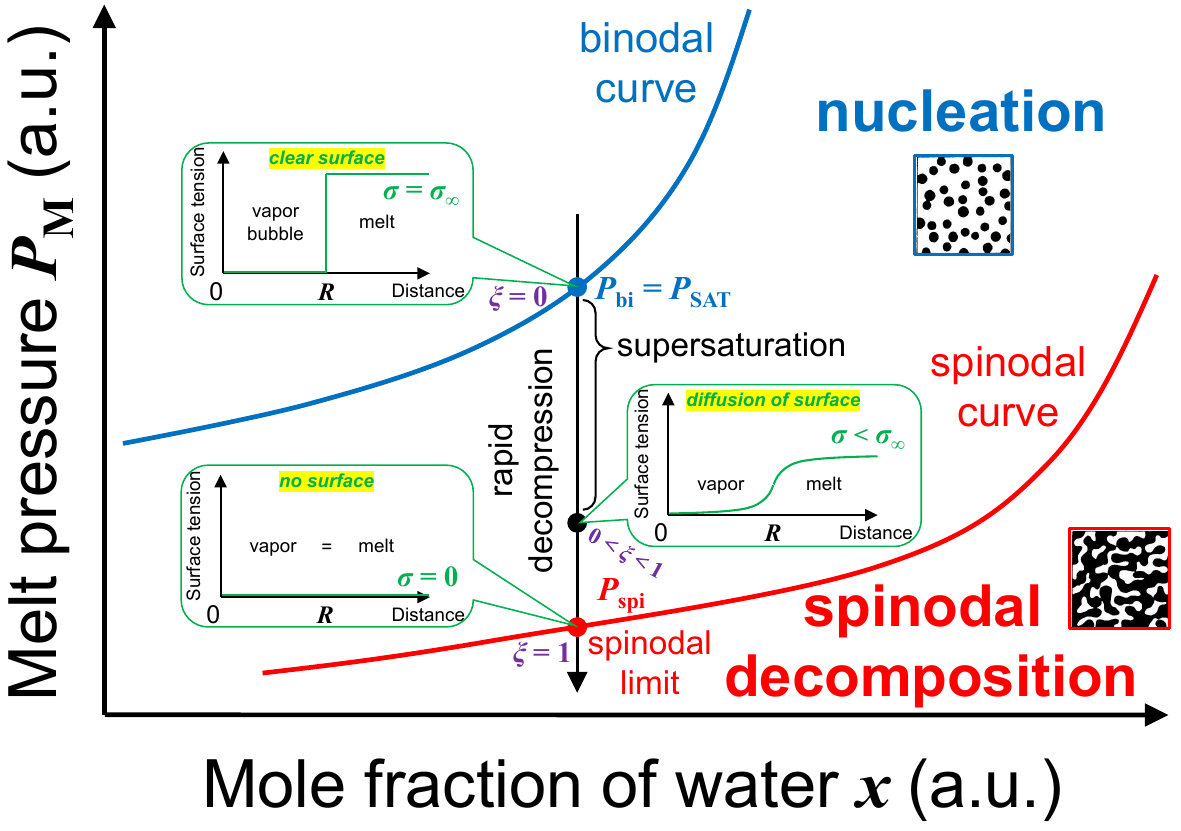}
    \caption{Schematic diagram illustrating the change in the “microscopic” surface tension $\sigma$ between the melt and homogeneous spherical bubble nucleus with decreasing pressure, based on nonclassical nucleation theory (Gonnermann and Gardner, 2013), in the phase diagram of the silicate--water system on the plane of mole fraction of water $x$--pressure $P$ at constant temperature. Both vertical and horizontal axes are in arbitrary units. The arrow indicates the rapid decompression path, representing a rapid decrease in melt pressure $P_\mathrm{M}$ and a rapid increase in the degree of supersaturation. See Eq. (\ref{xi}) for the definition of $\xi$, the index representing the degree of supersaturation. 
    Since the binodal curve coincides with the water solubility curve, the blue point on the curve, corresponding to the binodal pressure $P_\mathrm{bi}$, is equal to the saturation pressure $P_\mathrm{SAT}$, and the degree of supersaturation is zero ($\xi = 0$). At this point, nucleation occurs, and the interface between the melt and bubble nucleus is sharp and clear ($\sigma = \sigma_\infty$). 
    The black point, where decompression has further progressed, corresponds to a state with a certain degree of supersaturation ($0 < \xi < 1$). During nucleation at this point, the interface between the melt and bubble nucleus loses sharpness and diffuses ($\sigma < \sigma_\infty$). 
    The red point on the spinodal curve corresponds to the spinodal pressure $P_\mathrm{spi}$, where the degree of supersaturation reaches its maximum within the nucleation region ($\xi = 1$). At pressures below this point, spinodal decomposition occurs, in which phase separation begins without the formation of an interface between the melt and bubble nucleus ($\sigma = 0$).}
    \label{Fig6}
\end{figure}

Notations are listed in Table \ref{Table1}.
Based on the non-CNT, Gonnermann and Gardner (2013) considered a situation where the “microscopic” surface tension $\sigma$ depends on the degree of supersaturation: the greater the degree of supersaturation, the smaller $\sigma$ becomes.
This concept pertains to a non-equilibrium state; nevertheless, it can be interpreted using a chemical composition–pressure phase diagram at a constant temperature, which is thermodynamically determined. 
A schematic representation is shown in Fig. \ref{Fig6}. 
It should be noted that the binodal and spinodal curves assumed here are static, whereas the dynamic counterparts---taking into account kinetic effects such as differences in the viscoelastic properties between silicate melt and water vapor---are expected to shift to lower pressures (Nishiwaki, 2025).

The decompression path of magma on this phase diagram is determined by the competition between the timescales of supersaturation pressure accumulation due to decompression and its relaxation through the diffusion of water molecules into bubble nuclei (Nishiwaki, 2025). 
When the timescale of decompression is much longer than that of diffusion, i.e., when decompression occurs slowly, equilibrium degassing follows the binodal curve.
In contrast, when the timescale of decompression is much shorter than that of diffusion, the decompression path is expected to extend vertically downward, as indicated by the black arrow in the figure. 
In this case, the decrease in melt pressure $P_\mathrm{M}$ corresponds to an increase in the degree of supersaturation. 
In reality, even when the decompression timescale is $2.3 \times 10^3$ times longer than the diffusion timescale in decompression experiments, vesiculation is observed within the nucleation region (Hajimirza et al., 2019), making it extremely difficult to cross the spinodal curve without nucleation (Nishiwaki, 2025). 
Even so, in this study, I focus on rapid decompression to compare with the results of single-step decompression experiments discussed later, and I consider the path shown in the figure as a hypothetical decompression path. 
The rapid decompression path passes through the following three characteristic points as it progresses:

1) Blue point: Since the binodal curve corresponds to the water solubility curve, the pressure on the curve $P_\mathrm{bi}$ equals the saturation pressure $P_\mathrm{SAT}$ ($P_\mathrm{M} = P_\mathrm{bi} = P_\mathrm{SAT}$), and the degree of supersaturation is zero. If phase separation occurs at this point, nucleation takes place, and the interface between the melt and bubble nucleus is sharp and clear ($\sigma = \sigma_\infty$).

2) Black point: As the pressure drops through the nucleation region between the binodal and spinodal curves, the supersaturation increases. If phase separation occurs at this point, nucleation still occurs; however, the interface between the melt and bubble nucleus loses sharpness and diffuses, and $\sigma$ becomes lower than that in 1) ($\sigma < \sigma_\infty$).

3) Red point: When the spinodal curve is reached ($P_\mathrm{M}$ = spinodal pressure $P_\mathrm{spi}$), the degree of supersaturation reaches its maximum within the nucleation region. Below this pressure, spinodal decomposition occurs, in which phase separation begins without an interface between the melt and bubble nucleus ($\sigma = 0$).

Gonnermann and Gardner expressed this physical situation (the dependence of surface tension on the degree of supersaturation) using the following equation (identical to Eq. (17) in their paper):
\begin{equation}
\label{sigmanc}
\sigma = \sigma_\infty (1 - \xi^2)^{\frac{1}{3}} \qquad (0 \leq \xi \leq 1).
\end{equation}
Here, $\xi$ is an index representing the degree of supersaturation, and its definition is provided below in Eq. (\ref{xi}).
They used Eq. (\ref{sigmanc}) to model their decompression experiment results. They claimed that the modeling demonstrated that $\sigma$ strongly depends on the degree of supersaturation and that its value is smaller than $\sigma_\infty$. However, it should be noted that this logic appears self-contradictory. Their claim implies that the relation $\sigma < \sigma_\infty$ emerged as a modeling result, even though this relation had already been assumed in advance by defining the functional form of $\sigma$ within the nucleation region as shown above. 
Despite this logical inconsistency, their idea is innovative enough to warrant renewed attention. If their idea is correct, a relatively spinodal decomposition-like texture could emerge during decompression. For example, when supersaturation is large (i.e., when $\sigma$ is small), the distance between bubbles would follow an almost constant wavelength, even though the mechanism is nucleation.

Now, in Eq. (\ref{sigmanc}), the supersaturation index $\xi$ is defined as follows:
\begin{equation}
\label{xi}
\xi = \dfrac{P^*_\mathrm{B} - P_\mathrm{M}}{P^*_\mathrm{B} - P_\mathrm{spi}}, 
\end{equation}
where $P^*_\mathrm{B}$ is the internal pressure of a critical bubble nucleus. A bubble nucleus at the critical radius $R_\mathrm{c}$ satisfies both mechanical and chemical equilibrium with the surrounding melt.
The mechanical equilibrium is expressed by the Laplace equation:
\begin{equation}
\label{Laplace}
P^*_\mathrm{B} = P_\mathrm{M} + \dfrac{2 \sigma}{R_\mathrm{c}}.
\end{equation}
From these equations, it can be confirmed that on the binodal curve, $P_\mathrm{M} = P^*_\mathrm{B}$, i.e., $\xi = 0$, $\sigma = \sigma_\infty$, and $R_\mathrm{c}$ diverges to $\infty$, while on the spinodal curve, $P_\mathrm{M} = P_\mathrm{spi}$, i.e., $\xi = 1$ and $\sigma = 0$.
The chemical equilibrium is expressed as follows:
\begin{equation}
\label{P_g}
f (T, P^*_\mathrm{B}) = f (T, P_\mathrm{SAT}) \exp{ \left\{ \dfrac{\overline{V}_\mathrm{\ce{H2O}}}{k_\mathrm{B} T} (P_\mathrm{M} - P_\mathrm{SAT}) \right\} },
\end{equation}
where $f (T, P)$, the pure water vapor fugacity at this temperature and pressure, can be obtained from the real-gas equation of state for water (e.g., IAPWS95 by Wagner and Pru\ss, 2002). $\overline{V}_\mathrm{\ce{H2O}}$ is the partial molar volume of water in the melt and can be calculated from the empirical equation derived by Ochs and Lange (1999). This Eq. (\ref{P_g}) has been used to determine $P^*_\mathrm{B}$ since its introduction by Cluzel et al. (2008). The $P^*_\mathrm{B}$ values determined using this method, as summarized in Appendix table of Shea (2017), are relisted in Table \ref{Table2}. 

Gonnermann and Gardner (2013) assumed $P_\mathrm{spi}$ to be a hypothetical parameter and manually adjusted the numerator and denominator of the right-hand side of Eq. (\ref{xi}) to reproduce the experimental BND values. In contrast, $P_\mathrm{spi}$ can be computed according to Nishiwaki (2025), and $\sigma$ can be calculated directly from Eqs. (\ref{xi}) and (\ref{sigmanc}). 
It should be noted that, unlike Gonnermann and Gardner (2013), this method does not require the BND value of the decompression experiment product, as the value of $P_\mathrm{spi}$ is directly substituted. 

\section{Surface tension estimation by applying spinodal pressure into nonclassical nucleation theory}
\label{non-CNT}

In Table \ref{Table2}, I summarized the physical conditions of the single-step decompression (SSD) and homogeneous nucleation derived from previous decompression experiments (Gardner and Ketcham, 2011; Gardner, 2012; Gardner et al., 2013), as compiled in Shea’s (2017) Appendix table. Compared to continuous or multi-step decompression and heterogeneous nucleation, SSD and homogeneous nucleation are simpler processes. This simplicity makes them more amenable to the direct application of nucleation theory. In SSD, decompression to the final pressure is instantaneous; therefore, $P_\mathrm{M}$ can be regarded as equal to the final pressure. 

I calculated $\sigma$ for each experimental run.
To determine $\sigma$, it is first necessary to calculate the spinodal pressure $P_\mathrm{spi}$.
In Nishiwaki (2025), hydrous magma was modeled as a symmetrical regular solution composed of silicate melt and water vapor. Under this assumption, a relation was derived between the mole fractions of water $x$ on the binodal and spinodal curves at a given pressure $P$ (see Eq. (6) in the article). By applying this relation, along with the fact that the binodal curve coincides with the water solubility curve, I calculated $P_\mathrm{spi}$ for each experimental run.
For the temperature and pressure dependence of water solubility, the following empirical equations were used: 
for rhyolite, Liu et al. (2005);
for dacite, Eq. (2) in Gardner and Ketcham (2011);
for Na-phonolite: Eqs. (2) and (3) in Gardner (2012);
and for trachyte, basaltic andesite, and phonotephrite, Eq. (1) in Moore et al. (1998). 
As a result, very low values of $P_\mathrm{spi}$---no more than 3.4 MPa—--were obtained, as shown in Table \ref{Table2}.

\begin{figure}
    \centering
    \includegraphics[width=1.15\linewidth]{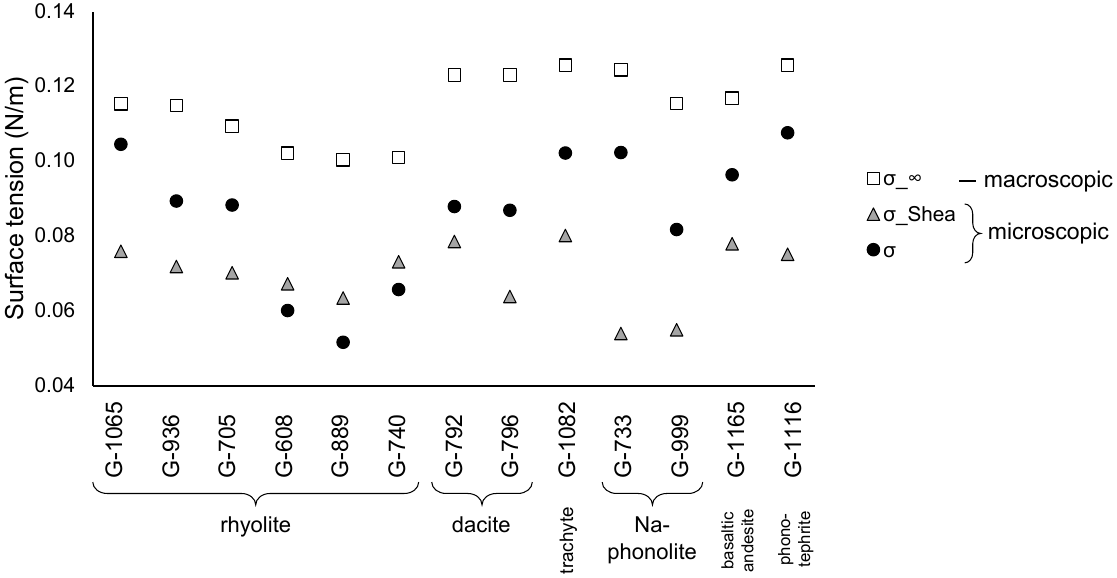}
    \caption{The “macroscopic” and “microscopic” surface tension values between the melt and bubble nucleus for various chemical composition and temperature conditions, as shown in Table \ref{Table2}. The meanings of $\sigma_\infty$, $\sigma_\mathrm{Shea}$, and $\sigma$ are given in the caption of Table \ref{Table2}.}
    \label{Fig7}
\end{figure}

The values of the microscopic surface tension $\sigma$ calculated in this study, along with $\sigma_\mathrm{Shea}$, obtained using the conventional method (i.e., inversion of BND using the CNT formula) in Shea (2017), and the macroscopic surface tension $\sigma_\infty$, calculated using the empirical formula proposed by Hajimirza et al. (2019), are presented in Table \ref{Table2} and Fig. \ref{Fig7}. 
The values of $\sigma$ and $\sigma_\mathrm{Shea}$ were notably lower than $\sigma_\infty$, and this trend is consistent with that reported in previous studies (e.g., Hamada et al., 2010). 
However, although many of the $\sigma$ values were higher than $\sigma_\mathrm{Shea}$, several were lower, and the variation among them was quite substantial.
In addition, among the $\sigma$ values listed in Table \ref{Table2}, two datasets that were also evaluated by Gonnermann and Gardner (2013) using their method (denoted as $\sigma_\mathrm{GG}$) enable direct comparison. 
For G-608, ($\sigma$, $\sigma_\mathrm{GG}$) = (0.060, 0.080), and for G-889, ($\sigma$, $\sigma_\mathrm{GG}$) = (0.052, 0.079) (both in N/m), indicating that the values obtained in this study are markedly smaller.
Therefore, the validity of the non-CNT-based Eqs. (\ref{sigmanc}) and (\ref{xi}) proposed by Gonnermann and Gardner (2013) cannot be conclusively determined, nor can the dependence of $\sigma$ on supersaturation. 

Even so, it should be noted that these results were obtained by applying the value of $P_\mathrm{spi}$ under the symmetric regular solution approximation. 
As mentioned in the Appendix of Nishiwaki (2025), the Gibbs energy of the real system composed of anhydrous silicate melt and water vapor is likely to be asymmetric, and a precise determination of $P_\mathrm{spi}$ requires a detailed understanding of this asymmetry. 
Nishiwaki (2023) proposed a thermodynamic approach that could potentially resolve this issue, and once implimented, it would allow for more accurate determinations of both $P_\mathrm{spi}$ and $\sigma$.

\begin{landscape}
\begin{table}[h]
\centering
\begin{tabular}{lrlrrrrrrrrr}
\hline
Melt composition  
& \multicolumn{1}{l}{\begin{tabular}[c]{@{}l@{}}\ce{SiO2}\\ (wt\%)\end{tabular}} 
& Run\#  & \multicolumn{1}{l}{\begin{tabular}[c]{@{}l@{}}$T$\\ ($^{\circ}$C)\end{tabular}} 
& \multicolumn{1}{l}{\begin{tabular}[c]{@{}l@{}}$P_\mathrm{SAT}$\\ (MPa)\end{tabular}} 
& \multicolumn{1}{l}{\begin{tabular}[c]{@{}l@{}}$P^*_\mathrm{B}$\\ (MPa)\end{tabular}} 
& \multicolumn{1}{l}{\begin{tabular}[c]{@{}l@{}}$P_\mathrm{M}$\\ (MPa)\end{tabular}} 
& \multicolumn{1}{l}{\begin{tabular}[c]{@{}l@{}}$P_\mathrm{spi}$\\ (MPa)\end{tabular}} 
& \multicolumn{1}{l}{\begin{tabular}[c]{@{}l@{}}$\xi$\\ (no unit)\end{tabular}}
& \multicolumn{1}{l}{\begin{tabular}[c]{@{}l@{}}$\sigma$\\ (N/m)\end{tabular}} 
& \multicolumn{1}{l}{\begin{tabular}[c]{@{}l@{}}$\sigma_\mathrm{Shea}$\\ (N/m)\end{tabular}} 
& \multicolumn{1}{l}{\begin{tabular}[c]{@{}l@{}}$\sigma_\infty$\\ (N/m)\end{tabular}} \\ 
\hline
rhyolite          & 76.53                                                                     & G-1065$^\mathrm{c}$ & 1150                                                                            & 190.0                                                                                & 152.6                                                                                & 75.5 & 0.25                                                                              & 0.506                     & 0.105                                                                        & 0.076                                                                                      & 0.115                                                                               \\
rhyolite          & 76.53                                                                     & G-936$^\mathrm{a}$  & 1085                                                                            & 172.0                                                                                & 139.1                                                                                & 38.0 & 0.13                                                                              & 0.727                     & 0.089                                                                        & 0.072                                                                                      & 0.115                                                                               \\
rhyolite          & 76.53                                                                     & G-705$^\mathrm{a}$  & 975                                                                             & 162.0                                                                                & 128.1                                                                                & 40.0 & 0.11                                                                              & 0.688                     & 0.088                                                                        & 0.070                                                                                      & 0.109                                                                               \\
rhyolite          & 76.53                                                                     & G-608$^\mathrm{a}$  & 875                                                                             & 161.0                                                                                & 124.3                                                                                & 13.5 & 0.14                                                                              & 0.892                     & 0.060                                                                        & 0.067                                                                                      & 0.102                                                                               \\
rhyolite          & 76.53                                                                     & G-889$^\mathrm{a}$  & 825                                                                             & 154.0                                                                                & 119.1                                                                                & 8.5 & 0.12                                                                               & 0.930                     & 0.052                                                                        & 0.064                                                                                      & 0.100                                                                               \\
rhyolite          & 76.53                                                                     & G-740$^\mathrm{a}$  & 775                                                                             & 140.0                                                                                & 100.3                                                                                & 15.0 & 0.08                                                                              & 0.851                     & 0.066                                                                        & 0.073                                                                                      & 0.101                                                                               \\
dacite            & 69.85                                                                     & G-792$^\mathrm{c}$  & 1150                                                                            & 160.0                                                                                & 128.3                                                                                & 26.5 & 0.54                                                                              & 0.797                     & 0.088                                                                        & 0.079                                                                                      & 0.123                                                                               \\
dacite            & 66.93                                                                     & G-796$^\mathrm{a}$  & 1150                                                                            & 160.0                                                                                & 136.0                                                                                & 27.0 & 0.54                                                                              & 0.805                     & 0.087                                                                        & 0.064                                                                                      & 0.123                                                                               \\
trachyte          & 62.57                                                                     & G-1082$^\mathrm{c}$ & 1150                                                                            & 151.5                                                                                & 121.0                                                                                & 39.0 & 0.09                                                                              & 0.678                     & 0.102                                                                        & 0.080                                                                                      & 0.126                                                                               \\
Na-phonolite      & 61.47                                                                     & G-733$^\mathrm{b}$  & 1150                                                                            & 155.0                                                                                & 137.0                                                                                & 48.0 & 3.4                                                                              & 0.666                     & 0.102                                                                        & 0.054                                                                                      & 0.125                                                                               \\
Na-phonolite      & 61.47                                                                     & G-999$^\mathrm{b}$  & 875                                                                             & 121.0                                                                                & 101.0                                                                                & 21.0 & 1.4                                                                              & 0.803                     & 0.082                                                                        & 0.055                                                                                      & 0.115                                                                               \\
basaltic andesite & 54.12                                                                     & G-1165$^\mathrm{c}$ & 1200                                                                            & 200.0                                                                                & 161.0                                                                                & 54.5 & 0.29                                                                              & 0.663                     & 0.096                                                                        & 0.078                                                                                      & 0.117                                                                               \\
phonotephrite     & 51.13                                                                     & G-1116$^\mathrm{c}$ & 1150                                                                            & 151.0                                                                                & 123.0                                                                                & 48.0 & 0.04                                                                              & 0.610                     & 0.108                                                                        & 0.075                                                                                      & 0.126                                                                               \\ \hline
\end{tabular}
\caption{\noindent {\footnotesize 
Surface tension values between the bubble nucleus and the surrounding melt correspond to each run of the previous single-step decompression (SSD) experiments for various chemical compositions and temperature conditions. All runs are those in which homogeneous nucleation was observed to occur and were extracted from Shea's (2017) Appendix table. The corresponding references (a: Gardner and Ketcham, 2011; b: Gardner, 2012; c: Gardner et al., 2013) are indicated at the top right of the run number Run\#.
$P_\mathrm{SAT}$: the saturation pressure; 
$P^*_\mathrm{B}$: the internal pressure of the critical bubble nucleus, calculated by Shea (2017) using Eq. (\ref{P_g}); 
$P_\mathrm{M}$: the melt pressure (equal to the final pressure of decompression); 
$P_\mathrm{spi}$: the spinodal pressure; 
$\xi$: the index representing the degree of supersaturation defined by Eq. (\ref{xi}); 
$\sigma$: the microscopic surface tension between the melt and bubble nucleus, obtained in this study to confirm the validity of the supersaturation dependence equation originally discussed in Gonnermann and Gardner (2013); 
$\sigma_\mathrm{Shea}$: the microscopic surface tension obtained by the conventional method---the inversion of BND using the classical nucleation theory (CNT) formula---in Shea (2017); 
$\sigma_\infty$: the macroscopic surface tension at the flat melt--vapor interface, calculated from the empirical formula Eq. (1) in Hajimirza et al. (2019).}}
\label{Table2} 
\end{table}
\end{landscape}

\section{Future directions in microscopic surface tension research}

Many BND data from decompression experiments significantly deviate from the values predicted by BND decompression rate meter (Fig. 4 in Fiege and Cichy, 2015). Moreover, the BND decompression rate meter tends to overestimate decompression rates more than other methods (Fig. 5 in Cassidy et al., 2018). These situations may be attributed to an inadequate estimation of $\sigma$.

The dependence of $\sigma$ on supersaturation was discussed in Section \ref{GG13} and \ref{non-CNT}. In addition, Hajimirza et al. (2019) discussed the dependence of $\sigma$ on another factor: the radius of a bubble nucleus, $R$. This dependence is known as the Tolman correction (Tolman, 1949) and is based on the following consideration: when the bubble size is very small (i.e., the interface curvature is large), as in nucleation, this geometric effect on molecular interactions results in the relation that $\sigma$ is lower than $\sigma_\infty$. The Tolman correction is expressed as follows:
\begin{equation}
\label{Tolman}
\dfrac{\sigma}{\sigma_\infty} = \dfrac{1}{1 + 2 \delta_\mathrm{T}/R},
\end{equation}
where $\delta_\mathrm{T}$ is a length scale called the Tolman length. By applying this theory to the results of decompression experiments, Gardner et al. (2023) found that $R$ and $\sigma$ increase under conditions of strong supersaturation. This finding contradicts the supersaturation dependence described by Gonnermann and Gardner (2013), and this inconsistency remains unresolved.

In light of the current situation, I propose to examine the integration of the dependence of surface tension $\sigma$ on both supersaturation and bubble nucleus radius. As stated by Gardner et al. (2023), a constant Tolman length $\delta_\mathrm{T}$, the assumption adopted in Hajimirza et al. (2019), is only valid for small deviations from the thermodynamic equilibrium and is likely to break down as nucleation proceeds at a higher degree of supersaturation (Joswiak et al., 2013). In other words, $\delta_\mathrm{T}$ appears to depend on the degree of supersaturation. This implies that the dependence of $\sigma$ on supersaturation and on bubble nucleus radius, which were considered separately, are interrelated. As shown in Eqs. (\ref{xi}) and (\ref{Laplace}), the bubble radius is determined by the degree of supersaturation; therefore, these two factors should not be considered separately. As mentioned in Section 3.6.1 of Toramaru (2022), a higher-order Tolman correction (e.g., Schmelzer and Baidakov, 2016; Tanaka et al., 2016) should be applied, since Eq. (\ref{Tolman}) currently used is only a first-order approximation.

Furthermore, there is room for improvement in the conventional method of estimating $\sigma$: the inversion of BND using the CNT formula.
I propose utilizing high-temporal-resolution data obtained from decompression experiments. Specifically, monitoring the time evolution of BND in single-step decompression (SSD) experiments---by varying the time between decompression and quenching/recovery of the sample---could provide valuable insight. In this way, the time-dependent variation of $J$ as a process of resolving the high supersaturation caused by rapid decompression could be revealed with high precision. In this regard, the work by Hajimirza et al. (2022) using rhyolitic melt represents the most detailed investigation of BND at each time point after decompression and has yielded the most accurate $J$ values reported to date. Their experimental results would therefore provide the most reliable $\sigma$ values.

Of course, the discrepancy between the theoretical and experimental BND values may also arise from factors other than surface tension. Preuss et al. (2016) mentioned that the diffusion coefficient of water should not be that of the total \ce{H2O}, but rather that of \ce{H2O}$_\mathrm{m}$, which is an order of magnitude larger than that of \ce{H2O}. This consideration would reduce the theoretical BND by an order of magnitude. Nishiwaki and Toramaru (2019) noted that high melt viscosity suppresses nucleation. The BND predicted by Toramaru’s (1995) viscosity-controlled regime may therefore be overestimated. Hence, it is necessary to carefully evaluate the discrepancy between the theoretical and experimental BND values by revisiting the fundamental physical processes of magma vesiculation, reexamining how each physical property contributes to nucleation, and reassigning appropriate values to the associated physical parameters.

\section{Conclusions}
\label{Conclusions}
I estimated the “microscopic” surface tension between the melt and bubble nucleus for previous decompression experiments by substituting the spinodal pressure into the equation describing the supersaturation dependence of surface tension, as proposed in the nonclassical nucleation theory by Gonnermann and Gardner (2013). However, the surface tension values obtained through this method exhibited significantly greater variability than those derived using the conventional approach (i.e., inversion of bubble number density using the CNT formula). This result suggests that the application of the nonclassical formula to magma vesiculation remains inconclusive at present.  
The discrepancy between theoretical and experimental BND values may stem from inaccuracies in surface tension estimation. Resolving this issue will require an integrated understanding of the dependence of surface tension on both supersaturation and bubble radius, combined with high–temporal-resolution experimental data to constrain nucleation processes more precisely.

\appendix
\section{Equation of the nucleation rate in classical nucleation theory (CNT)}
\label{Appendix}
The equation for $J$ in CNT (Hirth et al., 1970) is as follows:
\begin{equation}
\label{J}
J = \dfrac{2 n_0^2 D_\mathrm{\ce{H2O}} \overline{V}_\mathrm{\ce{H2O}}}{a_0} \sqrt{\dfrac{\sigma}{k_\mathrm{B} T}} \exp \left\{ - \dfrac{16 \pi \sigma^3}{3 k_\mathrm{B} T} (P^*_\mathrm{B} - P_\mathrm{M}) \right\},
\end{equation}
where $n_0$ is the number of water molecules per unit volume, $D_\mathrm{\ce{H2O}}$ is the diffusivity of total water in the melt, $\overline{V}_\mathrm{\ce{H2O}}$ is the partial molar volume of water in the melt, $a_0$ is the average distance between water molecules in the melt, $\sigma$ is the “microscopic” surface tension between the melt and the bubble nucleus, $k_\mathrm{B}$ is the Boltzmann’s constant, $T$ is the temperature, $P^*_\mathrm{B}$ is the internal pressure of the critical bubble nucleus, and $P_\mathrm{M}$ is the pressure of the melt.


\section*{Acknowledgments}
This work was supported by JSPS KAKENHI Grant Number JP23K19069.

\section*{References}
Allabar, A., and Nowak, M., 2018. Message in a bottle: Spontaneous phase separation of hydrous Vesuvius melt even at low decompression rates. \textit{Earth Planet. Sci. Lett.} \textbf{501}, 192--201. \url{https://doi.org/10.1016/j.epsl.2018.08.047}






Bagdassarov, N., Dorfman, A., and Dingwell, D. B., 2000. Effect of alkalis, phosphorus, and water on the surface tension of haplogranite melt. \textit{Am. Mineral.} \textbf{85} (1), 33--40. \url{https://doi.org/10.2138/am-2000-0105}





Cassidy, M., Manga, M., Cashman, K., and Bachmann, O., 2018. Controls on explosive-effusive volcanic eruption styles. \textit{Nat. Commun.} \textbf{9} (1), 1--16. \url{https://doi.org/10.1038/s41467-018-05293-3}




Cluzel, N., Laporte, D., Provost, A., and Kannewischer, I., 2008. Kinetics of heterogeneous bubble nucleation in rhyolitic melts: implications for the number density of bubbles in volcanic conduits and for pumice textures. \textit{Contrib. Mineral. Petrol.} \textbf{156} (6), 745--763. \url{https://doi.org/10.1007/s00410-008-0313-1}



Fiege, A., and Cichy, S. B., 2015. Experimental constraints on bubble formation and growth during magma ascent: A review. \textit{Am. Mineral.} \textbf{100} (11--12), 2426--2442. \url{https://doi.org/10.2138/am-2015-5296}

Gardner, J. E., 2012. Surface tension and bubble nucleation in phonolite magmas. \textit{Geochim. Cosmochim. Acta} \textbf{76}, 93--102. \url{https://doi.org/10.1016/j.gca.2011.10.017}

Gardner, J. E., and Ketcham, R. A., 2011. Bubble nucleation in rhyolite and dacite melts: temperature dependence of surface tension. \textit{Contrib. Mineral. Petrol.} \textbf{162} (5), 929--943. \url{https://doi.org/10.1007/s00410-011-0632-5}


Gardner, J. E., Ketcham, R. A., and Moore, G., 2013. Surface tension of hydrous silicate melts: Constraints on the impact of melt composition. \textit{J. Volcanol. Geotherm. Res.} \textbf{267}, 68--74. \url{https://doi.org/10.1016/j.jvolgeores.2013.09.007}

Gardner, J. E., Wadsworth, F. B., Carley, T. L., Llewellin, E. W., Kusumaatmaja, H., and Sahagian, D., 2023. Bubble formation in magma. \textit{Annu. Rev. Earth Planet. Sci.} \textbf{51}, 131--154. \url{https://doi.org/10.1146/annurev-earth-031621-080308}


Gonnermann, H. M., and Gardner, J. E., 2013. Homogeneous bubble nucleation in rhyolitic melt: Experiments and nonclassical theory. \textit{Geochem. Geophys. Geosyst.} \textbf{14} (11), 4758--4773. \url{https://doi.org/10.1002/ggge.20281}



Hajimirza, S., Gonnermann, H. M., Gardner, J. E., and Giachetti, T., 2019. Predicting homogeneous bubble nucleation in rhyolite. \textit{J. Geophys. Res. Solid Earth} \textbf{124} (3), 2395--2416. \url{https://doi.org/10.1029/2018JB015891}

Hajimirza, S., Gardner, J. E., and Gonnermann, H. M., 2022. Experimental demonstration of continuous bubble nucleation in rhyolite. \textit{J. Volcanol. Geotherm. Res.} \textbf{421}, 107417. \url{https://doi.org/10.1016/j.jvolgeores.2021.107417}

Hamada, M., Laporte, D., Cluzel, N., Koga, K. T., and Kawamoto, T., 2010. Simulating bubble number density of rhyolitic pumices from Plinian eruptions: constraints from fast decompression experiments. \textit{Bull. Volcanol.} \textbf{72} (6), 735--746. \url{https://doi.org/10.1007/s00445-010-0353-z}


Hirth, J. P., Pound, G. M., and St Pierre, G. R., 1970. Bubble nucleation. \textit{Metall. Trans.} \textbf{1} (4), 939--945. \url{https://doi.org/10.1007/BF02811776}





Jones, T. J., Cashman, K. V., Liu, E. J., Rust, A. C., and Scheu, B., 2022. Magma fragmentation: a perspective on emerging topics and future directions. \textit{Bull. Volcanol.} \textbf{84} (5), 45. \url{https://doi.org/10.1007/s00445-022-01555-7}

Joswiak, M. N., Duff, N., Doherty, M. F., and Peters, B., 2013. Size-dependent surface free energy and Tolman-corrected droplet nucleation of TIP4P/2005 water. \textit{J. Phys. Chem. Lett.} \textbf{4} (24), 4267--4272. \url{https://doi.org/10.1021/jz402226p}


Kelton, K. F., and Greer, A. L., 2010. Chapter 4 - Beyond the classical theory. In: Nucleation in condensed matter: Applications in materials and biology. Pergamon Materials Series \textbf{15}, 85--123. \url{https://doi.org/10.1016/S1470-1804(09)01504-1}




Liu, Y., Zhang, Y., and Behrens, H., 2005. Solubility of \ce{H2O} in rhyolitic melts at low pressures and a new empirical model for mixed \ce{H2O}--\ce{CO2} solubility in rhyolitic melts. \textit{J. Volcanol. Geotherm. Res.} \textbf{143} (1--3), 219--235. \url{https://doi.org/10.1016/j.jvolgeores.2004.09.019}





Maruishi, T., and Toramaru, A., 2024. Rapid coalescence of bubbles driven by buoyancy force: Implication for slug formation in basaltic eruptions. \textit{J. Geophys. Res.: Solid Earth} \textbf{129} (11), e2024JB029130. \url{https://doi.org/10.1029/2024JB029130}


Moore, G., Vennemann, T., and Carmichael, I. S. E., 1998. An empirical model for the solubility of \ce{H2O} in magmas to 3 kilobars. \textit{Am. Mineral.} \textbf{83} (1--2), 36--42. \url{https://doi.org/10.2138/am-1998-1-203}

Mourtada-Bonnefoi, C. C., and Laporte, D., 2004. Kinetics of bubble nucleation in a rhyolitic melt: an experimental study of the effect of ascent rate. Earth Planet. Sci. Lett. 218 (3--4), 521--537. \url{https://doi.org/10.1016/S0012-821X(03)00684-8}


Murase, T., and McBirney, A. R., 1973. Properties of some common igneous rocks and their melts at high temperatures. \textit{Geol. Soc. Am. Bull.} \textbf{84} (11), 3563--3592. \url{https://doi.org/10.1130/0016-7606(1973)84<3563:POSCIR>2.0.CO;2}



Nishiwaki, M., 2023. Chemical-thermodynamic explorations on the dissolution of water in magma: Breaking of the ideal mixing model and estimations of temperature change with decompression-induced vesiculation. Doctoral Dissertation, Kyushu University. \url{https://catalog.lib.kyushu-u.ac.jp/opac_detail_md/?reqCode=frombib&lang=0&amode=MD823&opkey=B168673756684664&bibid=6787423&start=1&bbinfo_disp=0}

Nishiwaki, M., 2025. Can spinodal decomposition occur during decompression-induced vesiculation of magma?. Preprint available in EarthArXiv. \url{https://eartharxiv.org/repository/view/6898/}

Nishiwaki, M., and Toramaru, A., 2019. Inclusion of viscosity into classical homogeneous nucleation theory for water bubbles in silicate melts: Reexamination of bubble number density in ascending magmas. \textit{J. Geophys. Res. Solid Earth} \textbf{124} (8), 8250--8266. \url{https://doi.org/10.1029/2019JB017796}

Ochs III, F. A., and Lange, R. A., 1999. The density of hydrous magmatic liquids. Science 283 (5406), 1314--1317. \url{https://doi.org/10.1126/science.283.5406.1314}

Ohashi, M., Ichihara, M., Kennedy, B., and Gravley, D., 2021. Comparison of bubble shape model results with textural analysis: implications for the velocity profile across a volcanic conduit. \textit{J. Geophys. Res.: Solid Earth} \textbf{126} (6), e2021JB021841. \url{https://doi.org/10.1029/2021JB021841}


Preuss, O., Marxer, H., Ulmer, S., Wolf, J., and Nowak, M., 2016. Degassing of hydrous trachytic Campi Flegrei and phonolitic Vesuvius melts: Experimental limitations and chances to study homogeneous bubble nucleation. \textit{Am. Mineral.} \textbf{101} (4), 859--875. \url{https://doi.org/10.2138/am-2016-5480}



Sahagian, D., and Carley, T. L., 2020. Explosive volcanic eruptions and spinodal decomposition: A different approach to deciphering the tiny bubble paradox. \textit{Geochem. Geophys. Geosyst.} \textbf{21} (6), e2019GC008898. \url{https://doi.org/10.1029/2019GC008898}

Schmelzer, J. W., and Baidakov, V. G., 2016. Comment on “Simple improvements to classical bubble nucleation models”. \textit{Phys. Rev. E} \textbf{94} (2), 026801. \url{https://doi.org/10.1103/PhysRevE.94.026801}


Shea, T., 2017. Bubble nucleation in magmas: A dominantly heterogeneous process?. \textit{J. Volcanol. Geotherm. Res.} \textbf{343}, 155--170. \url{https://doi.org/10.1016/j.jvolgeores.2017.06.025}








Tanaka, K. K., Tanaka, H., Ang\'elil, R., and Diemand, J., 2016. Reply to “Comment on ‘Simple improvements to classical bubble nucleation models’”. \textit{Phys. Rev. E} \textbf{94} (2), 026802. \url{https://doi.org/10.1103/PhysRevE.94.026802}

Tolman, R. C., 1949. The effect of droplet size on surface tension. \textit{J. Chem. Phys.} \textbf{17} (3), 333--337. \url{https://doi.org/10.1063/1.1747247}


Toramaru, A., 1995. Numerical study of nucleation and growth of bubbles in viscous magmas. \textit{J. Geophys. Res. Solid Earth} \textbf{100} (B2), 1913--1931. \url{https://doi.org/10.1029/94JB02775}


Toramaru, A., 2006. BND (bubble number density) decompression rate meter for explosive volcanic eruptions. \textit{J. Volcanol. Geotherm. Res.} \textbf{154} (3--4), 303--316. \url{https://doi.org/10.1016/j.jvolgeores.2006.03.027}

Toramaru, A., 2022. Vesiculation and crystallization of magma: Fundamentals of volcanic eruption process, conditions for magma vesiculation. Springer Singapore. \url{https://doi.org/10.1007/978-981-16-4209-8}

\url{https://doi.org/10.1186/s40623-025-02146-4}



Wagner, W., and Pru\ss, A., 2002. The IAPWS formulation 1995 for the thermodynamic properties of ordinary water substance for general and scientific use. \textit{J. Phys. Chem. Ref. Data} \textbf{31} (2), 387--535. \url{https://doi.org/10.1063/1.1461829}



Zhang, Y., and Ni, H., 2010. Diffusion of H, C, and O components in silicate melts. \textit{Rev. Mineral. Geochem.} \textbf{72} (1), 171--225. \url{https://doi.org/10.2138/rmg.2010.72.5}

\section*{CRediT authorship contribution statement}
Mizuki Nishiwaki:  Conceptualization, Methodology, Formal analysis, Investigation, Writing -- Original Draft, Writing -- Review \& Editing, Visualization, Project administration, Funding acquisition.

\section*{Declaration of competing interest}
The author declares that he has no known competing financial interests or personal relationships that could have appeared to influence the work reported in this paper.

\section*{Data availability}
The author confirms that the data supporting the findings of this study are available within the article.

\end{document}